# Open Ended Microwave Oven for Packaging

K. I. Sinclair[1], T. Tilford[2], M. Y. P. Desmulliez[1], G. Goussetis[1], C. Bailey[2], K. Parrott[2] and A. J. Sangster[1]

1. MicroSystems Engineering Centre (MISEC)
School of Engineering & Physical Science
Heriot Watt University
Edinburgh, EH14 4AS United Kingdom.

2. Centre for Numerical Modelling and Process Analysis (CNMPA)
University Of Greenwich
London, SE10 9LS United Kingdom

*Abstract* - A novel open waveguide cavity resonator is presented for the combined variable frequency microwave curing of bumps, underfills and encapsulants, as well as the alignment of devices for fast flip-chip assembly, direct chip attach (DCA) or wafer-scale level packaging (WSLP). This technology achieves radio frequency (RF) curing of adhesives used in microelectronics, optoelectronics and medical devices with potential simultaneous micron-scale alignment accuracy and bonding of devices. In principle, the open oven cavity can be fitted directly onto a flip-chip or wafer scale bonder and, as such, will allow for the bonding of devices through localised heating thus reducing the risk to thermally sensitive devices. Variable frequency microwave (VFM) heating and curing of an idealised polymer load is numerically simulated using a multi-physics approach. Electro-magnetic fields within a novel open ended microwave oven developed for use in micro-electronics manufacturing applications are solved using a dedicated Yee scheme finite-difference time-domain (FDTD) solver. Temperature distribution, degree of cure and thermal stresses are analysed using an Unstructured Finite Volume method (UFVM) multi-physics package. The polymer load was meshed for thermophysical analysis, whilst the microwave cavity – encompassing the polymer load – was meshed for microwave irradiation. The two solution domains are linked using a cross-mapping routine. The principle of heating using the evanescent fringing fields within the open-end of the cavity is demonstrated. A closed loop feedback routine is established allowing the temperature within a lossy sample to be controlled. A distribution of the temperature within the lossy sample is obtained by using a thermal imaging camera.

## I. Introduction

Microwave radiation fundamentally accelerates the cure kinetics of polymer adhesives. It provides a route to deposit heat energy primarily into the polymer materials [1]. Therefore, microwave radiation can be used to minimise the temperature increase in the surrounding materials such as the substrate and die during the cure process. This is especially important for devices incorporating either low thermal budget materials or interfaces with large thermal coefficient mismatch. The concentration of heat into the polymer during the cure process promotes its adhesion properties, as the magnitude of residual stress will be low between the polymer and materials, to which it is being bonded.

Multi-mode waveguide cavities have been used for the curing thermosetting resins [2], [3]. The ability to couple high power into a cavity oven enables fast polymerisation and large temperature gradients within the epoxy resin [4]. A material within a waveguide cavity oven with lower loss characteristics will remain at a lower temperature compared to the higher loss epoxy materials due to the selective nature of microwave energy absorption. Fundamentally, this characteristic reduces the thermal expansion mismatch between carrier and die allowing curing to be conducted in the presence of thermally sensitive components. As a consequence, it is possible to increase [or a least *not decrease*] the lifetime of the product [5]. Convention thermal curing heats the substrate first before transferring heat into the polymer while microwave curing process heats the polymer directly due to the volumetric nature of the absorption mechanism [6].

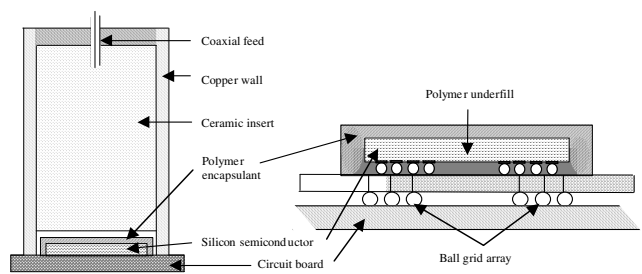

Fig. 1 Schematic showing layout of the cavity oven and the comparison between a 'conventional' assembly process and the proposed open-ended oven.

A maturing microwave curing technology is variable frequency microwave (VFM). Current VFM curing methods in electronic packaging bulk 'heats' the entire package or substrate [7]. For VFM, the source frequency is swept through a wide bandwidth, resulting in a scintillation of the model fields within the volume of the cavity resulting in a uniform field pattern. As such, a constant heating distribution can be obtained within homogenous materials [8]. In addition, sweeping the source frequency eliminates the likelihood of arcing thus allowing microwave heating of metals and semiconductor without damage due to electro-static discharge (ESD) [9]. Single frequency microwave (SFM) heating has been reported [10]. The source power can be pulased in order to avoid arcing.

Microwave heating has been successfully applied to heating glob-top and flip-chip underfill encapsulants [12]-[14], conductive and non-conductive epoxy adhesives [11], [15]-[16], polymer dielectrics [17]-[19], lead-free solder interconnects [20] and wafer bonding [21]. However, microwave heating is yet to be performed simultaneously with fine placement or alignment of the die or wafer onto the board. This lack of capability reduces the packaging throughput, as suggested by Fig. 1. The simultaneous fine placement and curing of the device into a larger assembly creates productivity gains by combining assembly, placement and bonding into a single processing step. The open-ended oven achieves curing by exposing the target package to the evanescent fields close to the 'open' exterior surface of a waveguide cavity resonator partially filled with a low loss dielectric [22]. While it is well known that evanescent fields have been used to generate electromagnetic coupling in a variety of microwave devices, the use of such fields for microwave heating within packaging assemblies is novel to the best of the authors' knowledge.

This paper presents a curing analysis of a polymer-silicon target sample using a novel multi-physics numerical solver and results obtained through experimental analysis of the system. Numerical results show VFM heating and curing of a idealised polymer material using the system to be feasible. Experimental results demonstrate heating within a lossy target sample placed at the boundary between the dielectric filling and the open region. The source power is adjusted using a closed feedback routine allowing the temperature to be controlled within the target sample. The thermal distribution within the sample is recorded using a thermal imaging camera and forms the basis of the feedback control.

## II. Open Ended Microwave Oven

The cavity oven takes the form of a rectangular waveguide partially filled with a low loss dielectric material. The target sample is placed at a location within the air filled section of the cavity. Fig. 1 shows a schematic representation of the cavity. Resonant modes can be trapped within the dielectric even if one end of the cavity is open due to the quasi-open circuit formed by the dielectric-air interface and the short-circuit at the copper wall. The fields within the air filled section are evanescent and attenuate exponentially with distance from the dielectric-air interface since the cut-off frequency of the air filled section of waveguide is less than the resonant frequency of the mode within the dielectric. Radiation losses can be minimised if the permittivity of the dielectric and the length of evanescent waveguide are chosen such that the evanescent fields are minimal at the open face. Power from the source is coupled into the cavity using either a probe (as in Fig. 1 and Fig. 3) located on the copper end wall or by slot aperture from waveguide. In order to take advantage of the continuity of the normal electric flux at the dielectric-air interface [23], the oven is operated using a transverse magnetic (TM) resonance.

A sample of lossy material, placed in the evanescent field, will experience ohmic heating if sufficient power is supplied. The absorbed power is primarily dependent on the magnitude of the average electric field within the sample volume and the dielectric loss tangent of the material itself [24]. In order to maximise the dielectric power loss within the sample, i.e. the heating potential, the target sample should have a large loss tangent relative to the dielectric filling. An efficient heating system would be one that has minimal loss within the dielectric filling and heavy dielectric losses within the target sample.

The prototype oven shown in Fig. 1 can operate with a single low order mode (for SFM) or, in principle, with multiple higher orders modes for VFM. Sweeping the source will couple a large number of modes (or combination of degenerate modes) producing scintillation of the field pattern within the transverse plane of the evanescent section. A discussion of the electromagnetic properties of the open-ended oven can be found in [23].

## III. Numerical Modelling

In order to accurately model the process of microwave curing a holistic approach must be taken. The process cannot be considered to be a sequence of discrete steps, but must be considered as a complex coupled system. Microwave energy is deposited into the dielectric materials, inducing heating. This in turn progresses a cure process and induces thermal stresses within the material. The system must be considered as a whole because each of these processes fundamentally influences the others. For example, the EM fields effect temperatures but are, in turn, affected by temperature and cure state through a change in the dielectric properties of the material. Thus, despite the very different thermal and electromagnetic timescales, there is a requirement to closely couple the computational electromagnetic, thermal, curing and stress analyses.

The numerical model used in this contribution is an extension of the approach developed for food processing applications [25]. The model comprises a FDTD electromagnetic solver coupled with an UFVM multi-physics package. Separate meshes are developed for electromagnetic and thermophysical solutions, with coupling implemented through an inter-domain cross mapping process. Overlapping solution domains are defined; the thermophysical analysis mesh is confined to the load material while the electromagnetic mesh occupies the entire oven domain (encompassing the thermophysical domain).

Electromagnetic analysis is carried in out in a rectangular domain encompassing the extent of the proposed cavity oven. A classical Yee scheme [26] has been implemented solving Maxwell's equations in the time domain with harmonic excitation on a tensor product computational mesh. In order to maintain numerical accuracy despite wavelength variation an automatic mesh generation algorithm has been employed.

Electromagnetic and thermophysical domains are interlinked through a cross mapping algorithm, based on a spatial sampling approach and capable of creating rapid conservative mappings between the meshes. Two sets of mappings are generated; firstly dielectric data is mapped from the thermophysical analysis domain into the FDTD

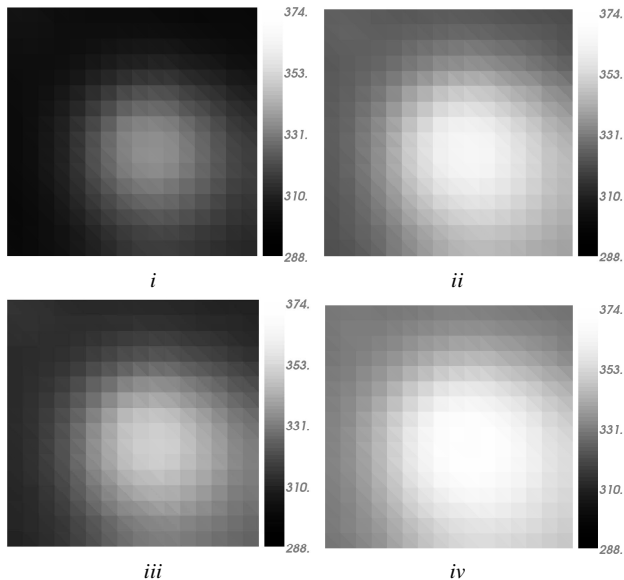

Fig. 2 (*i – iv*). Contour plots showing evolution of temperature (in degrees K) on upper surface of polymer load at consecutive 5 second intervals during curing process

domain prior to EM solution. Subsequently, power densities are mapped from the FDTD domain into the thermophysical analysis domain prior to solution of the thermophysical PDE's.

## IV. EXPERIMENTAL ANALYSIS

Experimental studies of the system have been conducted using a prototype system based on coupling of the $TM_{33}$ modes between 8Ghz and 12Ghz. Optimised coupling of the cavity over a wide bandwidth (required more VFM heating) has yet to be implemented within the design; As such, only SFM experimental results are presented within this paper.

A 25W travelling wave tube with protection circuitry and regulated power supply is connected to the cavity with a coaxial transmission line. A thermal imaging camera (X) is used to record the temperature distribution within the test sample and provides the basis for the control routine implemented using Labview [27]. A *temperature profile* can be specified allowing the temperature within the test sample to be controlled through automatic regulation of the source power using the feedback routine.

A cavity with dimensions 25.5mm x 25.5mm x 110mm was fabricated encompassing a low loss, ceramic dielectric material and is shown in Fig. 3. The dielectric filling has dimensions equal to 25.5mm x 25.5mm x 100mm with a relative permittivity at room temperature of 6 and tan $\delta$ of 0.0005 at 10GHz. The dielectric is recessed within the cavity at the open end, allowing for an air region of 10mm in length.

The test sample consists of a lead free solder paste placed between two circular coverslips of 15mm diameter with thickness of 0.15mm. The overall thickness of the sample can varied between 0.4mm and 0.6mm. This variation in test sample volume has insignificant impact on the electromagnetic performance of the cavity as the perturbation of the resoance frequency of the coulped mode is negligable. The complex permittivity of the lead-free solder was measured at room temperature using a dielectric probe (Agilent 85070E) and was found to have a relative permittivity of 4.6 and a tan $\delta$ of 0.6 at 10GHz. The cavity was excited by a coupling probe located centrally within the transverse plane on the copper end wall. The combination of parameters described above results in several $TM_{33}$ modes between 10GHz and 10.8GHz (a frequency range for which microwave measurement equipment is readily available). Fig. 4 displays the frequency response of the cavity over a bandwidth of 10GHz to 10.8GHz. The length of the probe is optimised to produce maximum coupling of the $TM_{335}$ mode resonant at 10.424GHz. $|S_{11}|$ = -13.2dB at this frequency indicating good impedance matching.

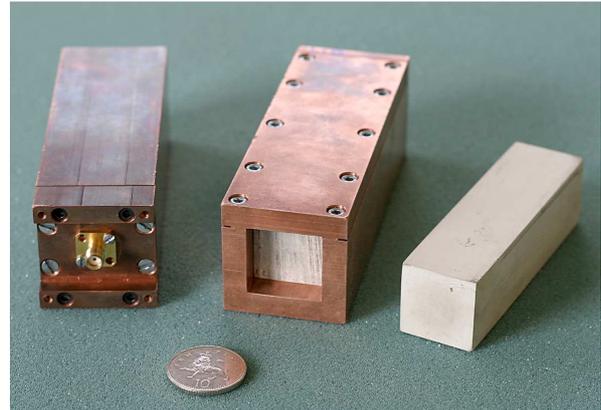

Fig. 3. Experimental cavity with dielectric shown on the right hand side.

A measurement of the cavity incorporating temperature feedback was conducted. The average temperature over the area of the coverslip is recorded using a thermal imaging camera. A target temperature versus time is plotted (red) over a period of 170s. The source power is automatically adjusted such that the temperature of the sample is controlled. Fig. 5 shows agreement between target and measured temperature within the lossy test sample thus demonstrating the validity of the control routine. The measurement was repeated with no sample present within the cavity. The temperature of the dielectric surface is plotted over the same period. Similar coupling is achieved for the $TM_{335}$ mode within this geometry allowing for comparative power levels to be delivered to the cavity. The surface temperature of the dielectric increases by 7°C after 170s. Therefore, the increase in the temperature within the test sample is primarily due to emersion within the evanescent field as opposed to thermal conduction from the dielectric filling material which is, in this example, negligible.

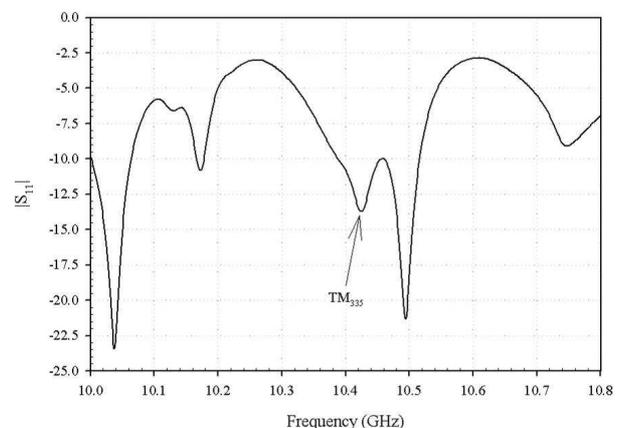

Fig. 4. Frequency response of the open-ended cavity incorporating low loss dielectric and lossy test sample.

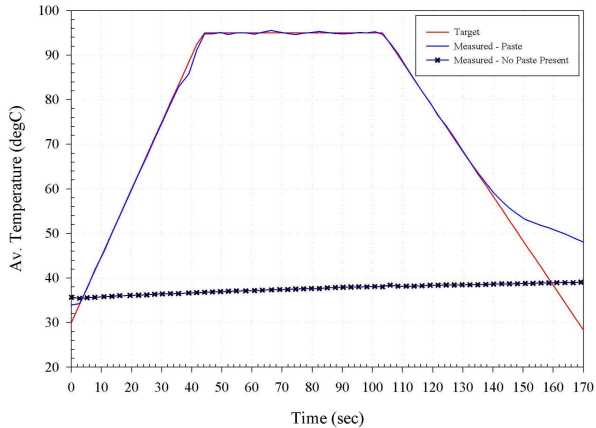

Fig. 5. Temperature versus time relationship showing the temperature observed within the test sample and at the surface of the dielectric when no sample present.

The temperature distribution within the test sample after 75s is plotted in Fig. 6. A uniform temperature distribution is displayed within the test sample. Although the cavity is optimised for the $TM_{335}$ mode, it is clear that the thermal distribution does not correlate with a 'pure' $TM_{3,3,5}$ mode pattern as anticipated. This is due to quasi-degeneracy and is attributable to the fact that the available Q-levels are insufficiently high, causing the modal resonances of neighboring modes to overlap. As such, unwanted resonant modes are be coupled into the cavity disturbing the idealised $TM_{335}$ field pattern. The effect of the quasi-degenracy can be reduced by lowering the dielectric loss of the dielectric material forming the resoanting region or by increasing the conductivity of the metal cavity walls. Quasi-degeneracy can also be suppressed by employing mode selective coupling involving the use of more than one excitation probe.

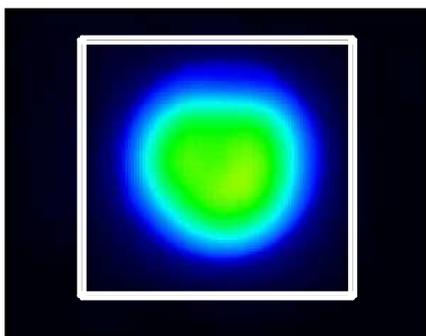

Fig. 6. Temperature distribution within target sample.

## V. CONCLUSION

This paper outlines the design and implementation of the open-ended waveguide cavity oven. A coupled numerical solver has been developed that allows the electromagnetic fields, temperature distribution, degree of cure and stress within the target package to be characterised. Numerical results demonstrate that, not only does the design of the cavity support heating using the evanescent fringing fields within a cutoff region of the cavity, but also that sufficient heating is obtained to facilitate curing of a polymer load. Experimental outputs demonstrate that control over the temperature within the load can be achieved through utilising a closed loop feedback routine. Finally, the experimental output supports the conclusion that the heating mechanism within the load occurs through ohmic losses due to interaction with the evanescent fringing field.

Further work includes the optimisation of the evanescent field through inclusion of a frequency selective surface (FSS) or layered dielectric configuration at the dielectric-air boundary and the characterisation of properties of microwave cured epoxies and encapsulant using the open-oven geometry.

ACKNOWLEDGEMENTS

This work was supported by the Innovative Electronics Manufacturing Centre (IeMRC), co-ordinated by Loughborough University, United Kingdom, through the project FAMOBS (FE/05/01/07). This work was also made possible through the funding of the project 3D-MINTEGRATION (EP/C534212/1) sponsored by the Engineering and Physical Sciences Research Council (EPSRC).